\documentclass{article}
\usepackage{spconf,amsmath,graphicx,amssymb,xcolor}
\usepackage[ruled]{algorithm2e}
\usepackage{epstopdf}
\usepackage{xcolor}

\def\v{{\mathbf v}}

\def\v#1{\mathbf{#1}}
\def\l#1{\mathcal{#1}}

\title{Semi-supervised Acoustic Event Detection based on tri-training}
\name{Bowen Shi\textsuperscript{1}, Ming Sun\textsuperscript{2}, Chieh-Chi Kao\textsuperscript{2}, Viktor Rozgic\textsuperscript{2}, Spyros Matsoukas\textsuperscript{2}, Chao Wang\textsuperscript{2}}
\address{\textsuperscript{1}Toyota Technological Institute at Chicago\\
\textsuperscript{2}Amazon\\
bshi@ttic.edu, \{mingsun,chiehchi,rozgicv,matsouka,wngcha\}@amazon.com}

\begin{document}
%
\maketitle
\begin{abstract}
This paper presents our work of training acoustic event detection (AED) models using unlabeled dataset. Recent acoustic event detectors are based on large-scale neural networks, which are typically trained with huge amounts of labeled data. 
Labels for acoustic events are expensive to obtain, and relevant acoustic event audios can be limited, especially for rare events. In this paper we leverage an Internet-scale unlabeled dataset with potential domain shift to improve the detection of acoustic events. 
Based on the classic tri-training approach, our proposed method shows accuracy improvement over both the supervised training baseline, and semi-supervised self-training set-up, in all pre-defined acoustic event detection tasks.
As our approach relies on ensemble models, we further show the improvements can be distilled to a single model via knowledge distillation, with the resulting single student model maintaining high accuracy of teacher ensemble models.
\end{abstract}

%
\begin{keywords}
acoustic event detection, tri-training, semi-supervised learning
\end{keywords}
\section{Introduction}
\label{sec:intro}
Acoustic event detection (AED) is the task of detecting whether certain events occur in an audio clip. It can be applied in many areas such as surveillance \cite{surveil_1, surveil_2}, and recommendation systems \cite{recommendation}. Conventionally AED has been addressed with automatic speech recognition techniques, e.g. with features such as mel-frequency cepstrum coefficients (MFCC) and classifiers based on hidden markov model (HMM).
In recent years, with the advances in speech recognition \cite{hinton12} and image recognition \cite{alexnet} as well as size increasing of datasets \cite{audioset}\cite{fb_data}, there are more deep learning based approaches applied to tackle AED tasks. For instance, the recently proposed Audioset \cite{audioset} comprises 1,789,621 10-second audio segments from a wide domain of 632 categories. 
Convolutional neural network (CNN) \cite{cnn_augment, many_cnns} or CNN-based approaches (e.g convolutional recurrent neural network \cite{rcnn, rcrnn}) are used and have shown improvements over traditional approaches. Though accuracy has been much improved in many AED tasks, state-of-the-art models often requires large number of labeled training data. Labeled data can be be quite limited under certain scenarios (e.g., for rare events \cite{aedr2}). The focus of this paper is on leveraging unlabeled audios to improve accuracy for AED.

Our main contributions include the following:
(1). We propose an ensemble method based on the classic tri-training that shows improvements in all acoustic events we investigate in a realistic semi-supervised setting (Internet-scale unlabeled dataset with domain shift)
(2). We show the improvements of the ensembled models can be distilled into a single model via knowledge distillation. As a result, there is no increase of computational costs during inference.

\section{Related work}
\label{sec:relate_work}

There has been a great volume of work on semi-supervised learning. A broad class of approaches contains feature learning with unlabeled data, based on generative models including variational autoencoders \cite{vae}, or generative adversarial networks \cite{gan1,gan2}. 
Another category of semi-supervised learning approaches is based on achieving certain smoothness effects with unlabeled data. For example, virtual adversarial training \cite{vat} relies on smoothing  model training with an regularization term based on adversarial direction. Those semi-supervised models are often evaluated in a 'simulated' setting by discarding many labels from an existing large labeled dataset, and they are sensitive to class distribution mismatch \cite{real_eval}.
Instead, our approach belongs to the family of bootstrapping methods, where models are often treated as a black box to assign pseudo labels on unlabeled data. 
Self-training is the simplest one of such category, which refers to retraining a model based on its own predictions on unlabeled data.
Despite its simplicity, self-training has been widely applied in practice. \cite{omni} proposed an approach based on self-training to visual structure prediction problems. For AED, self-training is employed to perform semi-supervised learning from Youtube audios \cite{aed_self}. 
Compared to previous efforts, our method is simpler but effective. The whole method can be directly built on audio features, e.g. log mel-filter bank energies (LFBEs), without involving complex data augmentation steps as in \cite{omni,mean_teacher}. Our experiments are placed in a realistic setting where unlabeled data come from Internet, and form a dissimilar distribution from the labeled dataset.


\section{Methods}
\label{sec:method}
In this section we describe the methods we use for semi-supervised learning. We focus on a multi-event classification setting.
Given an audio signal $I$ (e.g. LFBE), the task is to train a model $\v f$ to predict a multi-hot vector
 $\v y\in\{0, 1\}^C$ with $C$ being the size of event set $\l E$ 
and $y_c$ being a binary indicator whether event $c$ is present in $I$. Note the prediction $\v f(I)$ is not
a distribution over event set $\l E$ since multiple events can occur in $I$.
We denote $\l D_L=\{(I, \v y)\}$ as the labeled datset and $\l D_{UL}=\{I\}$ as the unlabeled dataset. 
In supervised setting, we train model $\v f$ using cross-entropy loss (see equation \ref{eq:sup_loss}), where $w_c$ is
the penalty of positive mis-classification of class $c$. $w_c$ is tuned to balance losses computed from positive 
and negative instances.

\begin{equation}
\centering
\label{eq:sup_loss}
L=-\displaystyle\sum_{(I, \v y)\in \l D_L}\displaystyle\sum_{c=1}^C\{w_c y_c\log f_c(I)+(1-y_c)\log(1-f_c(I))\}
\end{equation}


Self-training is a natural heuristic, which leverages a trained model to make predictions on unlabeled data and uses resulting pseudo labels to update the model
.  More formally, self-training consists of the following iterative process. A model $\v f$ is initially trained with minimizing loss defined in equation \ref{eq:sup_loss} with labeled data $\mathcal{D}_{L}$.  
At each iteration, we assign probability $\mathbf{p}(x) \in \mathbb{R}^C$ to every unlabeled example $x \in \l D_{UL}$ by applying model $\v f$. Top $k$ unlabeled data are selected for each class and added to the labeled dataset $L$ based on class score $m_c(\cdot)$, $\forall c\in \mathcal{C}$. 
Model $\v f$ is re-trained with labeled dataset $L^\prime$ augmented with $kC$ examples from $\mathcal{D}_{UL}$. Instead of directly setting a threshold for selecting data, we sort and select examples from $\mathcal{D}_{UL}$. As the model is applied on a different dataset $\mathcal{D}_{UL}$ with inevitable domain shifts at test time, relative order of confidence is more reliable than the absolute value of probabilty $\mathbf{p}$.



\textbf{Tri-training} One flaw of self-training is that the mistakes made by the model can be amplified by adding errorneous data. 
To avoid this, we can train multiple models and add data according to the agreement of those models.
In tri-training \cite{tri_training}, we first train three models $\v f_1$, $\v f_2$, $\v f_3$ independently. New data is added to train a particular model if other two models agrees on its label. In our multi-binary classification setting specifically, we train three different models by bootstrapping the training set.
Data $x$ is considered as a pseudo-label candidate for class $c$ of one model if its probability output by other two models are sufficiently high. We select top $k$ pseudo-labeled candidate data according to its score for each class (e.g. average probability of other two models) (see Algorithm \ref{alg:tri}). 
The data augmentation process is repeated for certain number of iterations. 
In classic tri-training, the final three models are ensembled during test time. However, in many real-world applications, the data distribution of unlabeled dataset can vary from labeled dataset, where the test set is from. Thus we ensemble models augmented through tri-training scheme, as well as the initial models trained with bootstrapped labeled data. As a results, there are in total 6 models ensembed, as shown at the end of Algorithm \ref{alg:tri}.

\vspace{-0.1in}
\begin{algorithm}[htp]
\SetAlgoLined
Initialize\;
\For{$i\in\{1, 2, 3\}$}{
$\l D^{0}_i= bootstrap(\l D_L)$\;
Train $\v f^0_i$ using eq.\ref{eq:sup_loss} with $\l D^{0}_i$\;
}
\For{$t\in\{1,..., T\}$}{
  \For{$i\in\{1, 2, 3\}$}{
    $\l D^t_i\leftarrow\emptyset$\;
    \For{$x\in\l D_{UL}$}{
      $\l P_c\leftarrow\emptyset$\;
      \For{$c\in\{1,...,C\}$}{
        \If{$f^{t-1}_{jc}(x)>\theta_c \land f^{t-1}_{hc}(x)>\theta_c$($j\neq h \neq i$)}{
          $\l P_c\leftarrow\l P_c\cup\{(x, \frac{f^{t-1}_{jc}(x)+f^{t-1}_{hc}(x)}{2})\}$\;
        }
        $\l D^t_i\leftarrow \l D^t_i\cup\text{top-k}(\l P_c)$\;
      }
    }
    Train $\v f^t_i$ using eq.\ref{eq:sup_loss} with $\displaystyle\bigcup_{t^\prime=0}^t D^{t^\prime}_i$\; 
  }
}
Ensemble $\v f^T_i$ and $\v f^0_i$($i\in\{1, 2, 3\}$)\;
\caption{\label{alg:tri} Ensemble-based tri-training}
\end{algorithm}

\vspace{-0.1in}
In our tri-training scheme, we need to rely on ensemble of models for test. In reality this can take a lot of time
and induce heavy memory and computation burdens, particularly for resource-constraint applications.
Thus we propose to ''trasnfer'' the ensembled models into a single
model with knowledge distillation. Specifically, we train a single model $\v f^s$ (student) to mimic its output distribution 
to the ensembled model $\v f^e$ by minimizing loss \ref{eq:kd_loss} adapted from the commonly used 
single-class classification setting \cite{kd}.

\begin{equation}
\label{eq:kd_loss}
\centering
\begin{split}
& L_{kd}=\displaystyle\sum_{(I, \v y)\in\l D_{L}}\{\alpha T^2 l(I, \v f^e(T))+(1-\alpha)l(I, \v y)\} \\
& l(I, \v y^\prime) = \displaystyle\sum_{c=1}^C\{w_c y^{\prime}_c\log f^s_c(I)+(1-y^{\prime}_c)\log(1-f^s_c(I))\} \\
& \v f^e(T) = \frac{1}{1+\exp(-\frac{\v z}{T})} \\
\end{split}
\end{equation}
, where $\v z$ is the logits of ensembled model. $T$ and $\alpha$ are hyperparameters  
controlling the softness of teacher logits $\v z$ and relative weight of distillation loss, respectively. For ensembling, we average the probabilities output by individual models and convert it back to logits. Note that only labeled dataset $\l D_L$ is used for training single student model.

\section{Experiments}
\label{sec:experiment}
\subsection{Experimental Setting}

\textbf{Data} The labeled dataset we use is a subset from Audioset \cite{audioset}. In particular, we select dog sound, baby crying and gunshots as the target events, which include both human and non-human vocals, as well as different durations of sound events. The three events included in Audioset amount to 13,460, 2,313 and 4,083 respectively, and we use all of them. All the three events are often considered rare events where number of labeled examples are quite limited in many real-world application scenarios. Note the class of dog contains any sounds produced by dog (e.g., barking, yipping), which makes the intra-class variation much bigger compared to other two events. 

In addition to the three events, we randomly selected 36,036 examples from all other audio clips in Audioset as negative samples. The negative vs. positive ratio is high especially for baby crying and gunshots ($>10$), which also aims to simulate the scarcity property of those rare events. We randomly split the whole subset for training ($70\%$), validation ($10\%$) and test ($20\%$). Additional efforts has been made to ensure the distribution of events roughly same across different sets. 

We use Amazon Instant Video (AIV) as our unlabeled dataset. The AIV data is a collection of audio parts of Amazon instant videos. To be consistent with Audioset, we split AIV audios into 10-second segments and the amount is 5,404,106 in total. Note the domain difference between AIV and Audioset can be large because AIV set is mainly media sounds (e.g. from films and TV shows) while the latter one contains many audio clips taken in real life.

\textbf{Implementation details} We first compute LFBE features for each audio clip. It is calculated with window size of 25 ms and hop size of 10 ms. The number of mel coefficients is 64, which gives us log-mel spectrogram feature of size $998\times 64$ for each audio clip. 
Features are further normalized by global cepstral mean and variance normalization (CMVN). 

We use DenseNet \cite{densenet} with 63 layers as our backbone model. The DenseNet we use contains 4 dense blocks with respectively 3, 6, 12 and 8 dense layers, where each layer is composed of batch normalization, ReLU, $1\times 1$ convolution, batch normalization, ReLU and $3\times 3$ convolution. The choice of model is based on dev performance under fully-supervised setting. As we have to run inference on large amount of unlabeled data, inference speed is also an important factor along with the accuracy. Our experimented models include ResNet \cite{resnet}, DenseNet and Conv-RNN \cite{conv_rnn1, conv_rnn2} with different layers, which are among the state-of-the-art models for acoustic event detection. According to our experiments, DenseNet-63 achieves highest accuracy  and also has relative small inference latency.

For ensemble-based tri-training, we pick the top 5,000 data for each class following algorithm \ref{alg:tri}. The number is tuned with dev set and will be analyzed in following sections. Model is re-trained from scratch when pseudo-labeled data are added. The tri-training process is repeated for one iteration. Here we did not observe further improvement by taking more iterations. For all experiments we use Adam optimizer with learning rate of 0.001 and batch size of 64.  We tuned penalty on positive loss ($w_c$) on dev set and found setting it to be the ratio between positive and negative examples of each class gives overall best results. This also prevents from tuning $w_c$ for every class under different settings.

\textbf{Evaluation Metric} We evaluate the performance of models based on area under curve (AUC) and equal error rate (EER) on detection error tradeoff (DET) curve (vertical: false negative rate (FNR), horizontal: false positive rate (FPR)). Performance is measured for individual events.

\textbf{Baselines} We compare ensemble based tri-training with the following two baselines. (1) Fully-supervised DenseNet-63, and (2) Self-training. Both baseline model training follows the same experimental setting as tri-training, but with a single DenseNet-63. 

\subsection{Results}
The results of different models are shown in table \ref{tab:main_result}. The proposed ensemble-based tri-training outperforms other semi-supervised learning approaches in all three tasks. Detailed analysis on the improvements will be presented in the following analysis section. In principle, semi-supervised approaches should be lower-bounded by the fully-supervised baseline.
But as we evaluate classes individually, it is possible to have degradation for some classes. For gunshots, all semi-supervised approaches improves over the supervised baseline. This may be related to the small domain discrepancy between unlabeled and labeled datasets for this particular event. We find many gunshot audioclips in Audioset are from multi-media source (e.g. video games), which is similar to the unlabeled AIV data. This shows semi-supervised learning helps especially when labeled and unlabeled data are from same domain. 

\begin{table}[ht]
\centering
\small
\begingroup
\setlength{\tabcolsep}{2pt}
\begin{tabular}{c||cccc|cccc}
 & \multicolumn{4}{c|}{AUC (\%)} &\multicolumn{4}{c}{EER (\%)} \\
Event & Sup$_s$ & Self$_s$ & Tri$_e$ & Tri-KD$_s$ & Sup$_s$ & Self$_s$ & Tri$_e$ & Tri-KD$_s$ \\ \hline
Dog & 4.32 & 4.42 & \bf 3.26 & 3.49 & 11.11 & 11.07 & \bf 9.29 & 9.80 \\
Baby-cry & 2.20 & 2.89 & \bf 1.42 & 1.69 & 6.56 & 7.34 & \bf 5.41 & 6.01 \\
Gunshots & 2.07 & 1.77 & \bf 1.31 & 1.51 & 6.41 & 5.78 & \bf 4.70 & 5.41 \\
\end{tabular}
\caption{\label{tab:main_result} Performance of models (on test set). Lower is better. Sup: fully-supervised baseline, Self: self-training,
Tri: ensemble-based tri-training, Tri-KD: distilled model from ensembled tri-training models,
$e$: Ensembled model, $s$: Single model}
\endgroup
\end{table}

The results of knowledge distillation trained DenseNet-63 (Tri-KD) with ensemble of tri-training models as teacher are also listed in table \ref{tab:main_result}. Though there is small degradation compared to the tri-training, it outperforms the other single models. The improvement shows that it is possible to distill the gain brought by using large amount of unlabeled data into a single supervised-trained model, so that there are no additional computational costs during inference.

\textbf{Ablation Study}
There are three factors contributing to the performance improvements for semi-supervised training: the scale of model, unlabeled data, ensemble of models. We further study the effects of these factors on the models and results are shown in table \ref{tab:tri_fac}. To avoid tuning on test set, all analysis are done on dev set. Compared to other approaches, tri-training has more diversity as three models are trained by bootstrapping the original training set. This increased scale brings improvements over the baseline even without any unlabeled data. 
Adding unlabeled data improves the performance in general despite small degradation of EER on baby crying. In binary classification on imbalanced data, adding pseudo-labeled data balances the training set and we observed model converged much faster compared to supervised baseline. The unapparent improvement on baby crying is related to the domain difference between the unlabeled and labeled datasets. We also observe that simple ensemble of models trained with and without unlabeled data can mitigate the side-effects of using unlabeled data brought by domain discrepancy.

\begin{table}[htp]
\centering
\small
\begingroup
\setlength{\tabcolsep}{2pt}
\begin{tabular}{c||cccc|cccc}
 & \multicolumn{4}{c|}{AUC (\%)} &\multicolumn{4}{c}{EER (\%)} \\
Event & Sup & +Ens & \begin{tabular}{@{}c@{}} +Ens\\+Data \end{tabular} & \begin{tabular}{@{}c@{}} +2xEns\\+Data \end{tabular} & Sup & +Ens & \begin{tabular}{@{}c@{}} +Ens\\+Data \end{tabular} & \begin{tabular}{@{}c@{}} +2xEns\\+Data \end{tabular} \\ \hline
Dog & 4.48 & 3.96 & 3.29 & \bf 3.28  & 11.06 & 9.81 & 9.09 & \bf 8.95\\
Baby-cry & 2.89  & 2.86 & 2.75 & \bf 2.57 & 8.30 & \bf 7.95 & 8.71 & 8.21\\
Gunshots & 2.46 & 1.53 & 1.39 & \bf 1.28 & 7.68 & 6.11 & 5.41 & \bf 5.22 \\
\end{tabular}
\caption{\label{tab:tri_fac} How different factors contributes to the performance of tri-training (on dev set). Lower is better. 
Sup: supervised baseline, +Ens: ensembled 3 models trained with only labeled data, +Ens+Data: ensembled tri-training models with unlabeled data, 2xEns+Data: ensembled tri-training models with and without unlabeled data.
Note that Sup in table \ref{tab:tri_fac} and Sup$_s$ in table \ref{tab:main_result}, 2xEns+Data in table \ref{tab:tri_fac} and Tri in table \ref{tab:main_result} refer to same approach.
}
\endgroup
\end{table}

\vspace{-0.1in}
\textbf{Varying amount of pseudo-labeled data} 
Number of pseudo-labeled data to add is an important hyper-parameter to tune. Table \ref{tab:tri_num} summarizes our analysis on this front.  Adding more pseudo-labeled data raises the  percentage of data with wrong labels in training set. We find that within certain range, the side-effects brought by the noisy data can be compensated by the data amount. Adding few data with high confidence is not as effective, because those are mainly "easy" data with which models are not guaranteed to be strengthened. 
Varying the data amount does not have as much impact on baby crying as other two events, which may be related to the domain shift of this particular event in unlabeled dataset.

\vspace{-0.1in}
\begin{table}[htp]
\centering
\small
\begingroup
\setlength{\tabcolsep}{5pt}
\begin{tabular}{c||ccc|ccc}
 & \multicolumn{3}{c|}{AUC (\%)} &\multicolumn{3}{c}{EER (\%)} \\
Event & 1k & 5k & 10k & 1k & 5k & 10k \\ \hline
Dog & 3.82 & \bf 3.28 & 4.02 & 10.02 & \bf 8.95 & 10.46 \\
Baby-cry & 2.69 & \bf 2.57 & 2.82 & \bf 8.20 & 8.21 & 8.73 \\
Gunshots & 2.09 & \bf 1.28 & 1.75 & 6.46 & \bf 5.22 & 6.16\\
\end{tabular}
\caption{\label{tab:tri_num} How number of pseudo-labeled data impact performance (of Tri in table \ref{tab:main_result}, on dev set). Lower is better. Our experimental results described earlier are based on 5k.}
\endgroup
\end{table}

\begin{figure}[htp]
\centering
\includegraphics[width=\linewidth]{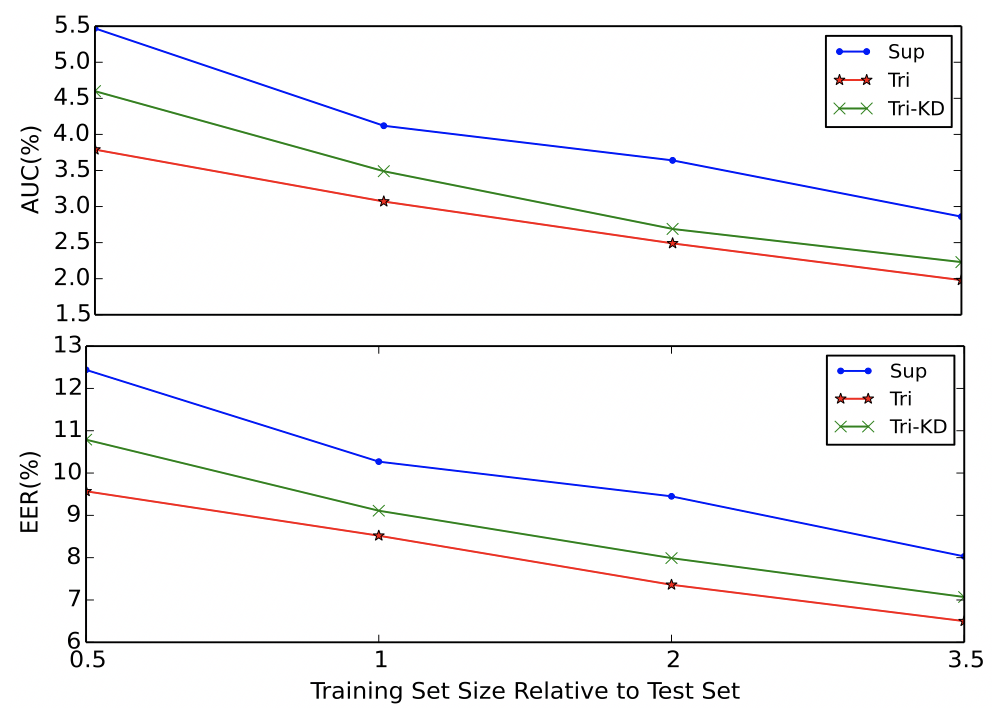}
\caption{\label{fig:variance} Average AUC (\%) and EER (\%) (on test) of three events with different training set to test set ratio. Training dataset is sampled according to each train vs. test ratio} 
\end{figure}

\vspace{-0.1in}
\textbf{Varying size of labeled training set}
In our default experimental setting, we have a relatively larger training set (ratio of train set to test set = 3.5). To see how the model performs with different size of training set, we reduced number of training data. Specifically we keep the same test and validation set and change the ratio between training and test set to \{0.5, 1.0, 2.0, 3.5\} 
(3.5 is the whole original training set). According to figure \ref{fig:variance},  our semi-supervised learning approach shows consistent gains with different size of training set, and using unlabeled data brings more gain when training set is relatively smaller. 

\vspace{-0.15in}
\section{Conclusions}
\label{sec:conclusions}
We investigate using large number of unlabeled data to improve acoustic event detection. Our proposed approach which is based on classic tri-training with ensembling shows consistent improvements over models trained with labeled data, as well as with self-training. In addition, we show that such improvements brought by the ensembled tri-training models can be distilled into a single model, which shows improved accuracy with same computational cost during inference.





\bibliographystyle{IEEEbib}
\bibliography{strings,refs}

\end{document}